\documentclass[prb,twocolumn,floatfix,showpacs,amssymb]{revtex4}
\usepackage{graphicx}
\usepackage{dcolumn}
\usepackage{bm}

\begin{document}

\title{Energy relaxation in disordered
charge and spin density waves}

\author{R. M\'elin$^{(1)}$\footnote{E-mail: melin@grenoble.cnrs.fr},
K. Biljakovi\'c$^{(2)}$ and J.C. Lasjaunias$^{(1)}$}
\affiliation{ 
$^{(1)}$ Centre de Recherches sur les Tr\`es Basses
Temp\'eratures (CRTBT)\footnote{U.P.R. 5001 du CNRS, Laboratoire conventionn\'e
avec l'Universit\'e Joseph Fourier},\\ Bo\^{\i}te Postale 166,
F-38042 Grenoble Cedex 9, France\\
$^{(2)}$ Institute of Physics,
Hr-10 001 Zagreb, P.O. Box 304, Croatia}

\begin{abstract}
We investigate collective effects
in the strong pinning model of disordered charge
and spin density waves (CDWs and SDWs) in connection with heat relaxation
experiments.
We discuss the classical and quantum limits that
contribute to two distinct contribution to the specific heat (a $C_v \sim
T^{-2}$ contribution and a $C_v \sim T^{\alpha}$ contribution respectively),
with two different types of disorder (strong pinning versus
substitutional impurities).
From the calculation of the two level system energy splitting distribution
in the classical limit
we find no slow relaxation
in the commensurate case and a broad spectrum of relaxation times
in the incommensurate case. In the commensurate case quantum effects
restore a non vanishing energy relaxation, and
generate stronger disorder
effects in incommensurate systems.
For substitutional disorder we obtain
Friedel oscillations
of bound states close to the Fermi energy. 
With negligible interchain couplings
this explains the power-law specific heat $C_v \sim T^{\alpha}$
observed in experiments on CDWs and SDWs
combined to the power-law susceptibility
$\chi(T)\sim T^{-1+\alpha}$ observed in the CDW o-TaS$_3$.
\end{abstract}


\maketitle

\section{Introduction}

There exist many examples of systems
showing slow relaxation and ageing: spin
glasses\cite{spin-glass1,spin-glass2}, disordered
dielectrics\cite{dielectrics1,dielectrics2,dielectrics3},
supercooled liquids\cite{glycerol}, etc ... Charge density
waves (CDWs) and spin density waves
(SDWs)\cite{manips-AsF6,manips-TaS3,manips-PF6-JPCM-I,manips-PF6} show
``interrupted ageing''\cite{trap-model}, meaning that 
there exists an upper bound $\tau_{\rm max}$ to the relaxation times.
The protocol of
ageing experiments in CDWs and SDWs is the following: the system
is equilibrated at the temperature $T$ (one
waits for a time longer than $\tau_{\rm max}$). At time $t=0$ the
temperature is changed from $T$ to $T+\Delta T$ where $\Delta T>0$
is very small compared to $T$. The temperature is kept constant
until the waiting time $t_w$ where it is brought back to $T$.
The heat flows between the CDW or SDW sample and the cold
source are recorded as a function of time. 
Ageing in the thermal response takes place even for
very small values of $\Delta T$. Since 
very small temperature variations
are applied in the experiment it is reasonable to
suppose that
the size of the correlated objects does not evolve in time
and that the thermal response is due solely to the variation
in the population of metastable states. This can be contrasted
with the coarsening dynamics where the size of correlated
domains increases with time.

In a recent work\cite{Melin-CDW} we applied the idea of
dynamical renormalization
group\cite{Dasgupta-Ma,FLDM}
to calculate the spectrum of relaxation times of a model of disordered
CDW or SDW\cite{Fukuyama,FLR,Larkin1,Larkin2,Abe,Rammal,Braso,Larkin3,Ov,revue},
including interactions among bisolitons.
A drawback of this approach\cite{Melin-CDW} is that we
supposed a coarsening dynamics following a quench from high
temperature, a situation that is not realized in experiments,
and we were not able to address the waiting time dependence of the
relaxation time spectra. One goal of the present article is to
address these issues that were left open in our previous work\cite{Melin-CDW},
and to put on a microscopic basis the random energy-like (REM-like) trap
model that was proposed in Ref.~\onlinecite{Melin-CDW},
and inspired from trap models developed for glasses and spin
glasses\cite{trap-model,Derrida}.

More specifically
we show here that heat relaxation experiments can be described by
assuming two types of defects 
(strong pinning and substitutional impurities),
corresponding to the ``classical''
limit where the CDW or SDW is viewed as a classical elastic medium
with bisolitons generated by strong pinning impurities
distributed
at random, and to the ``quantum'' limit where solitons due to
substitutional disorder
interact quantum mechanically by excitations of the gaped background.
The existence of two effects is in agreement with the experimental
observation that the low temperature out-of-equilibrium specific heat can be
decomposed into three contributions: (i) the $C_v \sim 1/T^2$ tail of a Schottky anomaly
at very low temperature (typically for $T \alt 100 \div 300$~mK;
the upper bound depends on the amplitude of the $1/T^2$ contribution);
(ii) a $C_v \sim T^{\alpha}$ power-law
specific heat with $\alpha \simeq 0.3 \div 1.2$ 
at intermediate temperatures ($0.1 \alt T \alt 1$~K);
and (iii) the ``trivial'' contribution of phonons $C_v \sim T^3$ at high
temperature ($T \agt 1$~K). By Schottky anomaly we mean that the equilibrium
specific heat
of a two-level system with energies $E_0$ and $E_0+\Delta E$ is given by
\begin{equation}
C_v(T)=\frac{(\Delta E)^2}{4 T^2}
\frac{1}{\cosh^2{(\Delta E/2 T)}}
,
\end{equation}
having a maximum (called a Schottky anomaly)
at $T_{\rm max} \simeq 0.416 \Delta E$. The specific heat is 
approximately equal to $C_v \simeq (\Delta E)^2/4 T^2$ in
the large temperature tail.

Following Ref.~\onlinecite{Ov},
the contribution (i) is interpreted in terms of two-level systems due to
strong pinning impurities. The contribution (ii)
is interpreted as midgap states interacting through Friedel oscillations.
Friedel oscillations of a single impurity were probed directly
by x-ray diffraction experiments in Ref.~\onlinecite{Pouget}.
Another evidence in favor of the coexistence of strong pinning and
substitutional impurities is that the CDW compound o-TaS$_3$ can be doped
by Nb, a substitutional impurity. This changes only the amplitude of the 
$C_v \sim T^\alpha$ contribution, but leaves unchanged the
$C_v \sim 1/T^2$
contribution\cite{Bilja-2003}, suggesting that the power-law
contribution is related to substitutional disorder. Even though
not affected by substitutional disorder we do not interpret 
the $C_v \sim 1/T^2$ as a property of the pure compound.
A nuclear hyperfine interaction can be excluded from the systematic
study of many different CDW compounds\cite{manips-AsF6}. We thus relate
the $C_v \sim 1/T^2$ contribution to strong pinning impurities\cite{Ov},
even though the microscopic nature of these impurities is not
well understood experimentally (see Ref.~\onlinecite{Dumas} for
a study of ESR spectroscopy in o-TaS$_3$).

The commensurate organic spin-Peierls compound
(TMTTF)$_2$PF$_6$
showing slow relaxation\cite{new-ref}
contrasts with the inorganic spin-Peierls
compound Cu$_{1-x}$Zn$_x$GeO$_3$\cite{CuGeO3-1,CuGeO3-2,CuGeO3-3,CuGeO3-4}
showing antiferromagnetic ordering.
We argue that the
difference
lies in the different nature of disorder.
Substitutional disorder relevant to Cu$_{1-x}$Zn$_x$GeO$_3$
is qualitatively different from strong pinning impurities in CDWs and SDWs. 
The spin-Peierls compound Cu$_{1-x}$Zn$_x$GeO$_3$ has a fast dynamics,
with a ``microscopic'' time presumably comparable to the one
of spin glasses ($\tau_0\simeq 10^{-12}$ sec) whereas in CDW and SDW
compounds we have $\tau_0\simeq 1$ sec for the thermally activated process.
This indicates that rather different mechanisms are at work,
identified here are substitutional or strong pinning disorder.
We generalize to the incommensurate case the model of substitutional
disorder introduced in Ref.~\onlinecite{FM}. In this model the solitons
are due to domain walls between two degenerate ground states since the
impurity site can be removed from the chain (see
Ref.~\onlinecite{FM} and section~\ref{sec:quantum}),
therefore leaving randomly distributed domain walls in the chain
from which the impurities sites have been removed.
The specificity of this model (as opposed to the strong pinning model) is that
there are no metastable states of bisolitons like in the strong
pinning model. 
The substitutional impurities do not contribute to the
slow dynamics of bisolitons in the strong pinning limit
but are expected to contribute to collective pinning of the phase of the
density wave. However we consider here temperatures much lower than the
glass transition temperature\cite{Hosseini,Bechgaard} so that
interactions among solitons are the only remaining
collective effects.
For substitutional disorder we find
interactions among solitons due to Friedel oscillations.
We obtain similarly to Ref.~\onlinecite{FM}
a power-law specific heat
$C_v(T) \sim T^{\alpha}$ and a susceptibility $\chi(T) \sim T^{-1+\alpha}$
in agreement with existing experiments on the CDW o-TaS$_3$\cite{Bilja-2003}.

The article is organized as follows. In section~\ref{sec:cl}
we investigate a classical model of collective effects in
a disordered CDW.
Quantum effects are investigated in 
section~\ref{sec:quantum}. Final remarks are given in
section~\ref{sec:conclu}.

\section{Classical limit (strong pinning impurities)}
\label{sec:cl}

\subsection{Hamiltonian}
Let us start with a classical model of disordered
CDW\cite{Fukuyama,FLR,Larkin1,Larkin2,Abe,Rammal,Braso,Larkin3,Ov,revue}.
To derive the 1D projection of the Hamiltonian of the phase of the
CDW in the mean field approximation we follow 
the recent review by Brazovskii and Nattermann\cite{revue}
and consider a system of coupled chains with a phase $\varphi_n(y)$
in chain $n$ ($y$ is the coordinate along the chain axis).
\begin{eqnarray}
\label{eq:H-n}
{\cal H} &=& 
\frac{\hbar v_F}{4\pi} \sum_n \int dy \left(
\frac{\partial \varphi_n(y)}{\partial y} \right)^2\\
&+& \sum_{n,m} w_{m,n} \int dy
\left[1-\cos{(\varphi_n(y)-\varphi_m(y))}\right]\\
&-&\sum_{n,i} V_i^{(n)}
\left[1-\cos{\left(Q y_i^{(n)}+
\varphi_n(y_i^{(n)}) \right)}\right]
,
\end{eqnarray}
where the sum in the last term runs over all impurities,
$v_F$ is the Fermi velocity along the chain axis,
$w_{m,n}$ corresponds to the commensurate energy or interchain
coupling, $V_i^{(n)}$ is the pinning energy of the impurity
number $i$ in chain $n$, and $Q=2 k_F$ is the wave vector of the
CDW. Assuming dilute impurities we suppose
that the chain $n=0$ with $\varphi_0(x)\equiv \varphi(x)$
is coupled to neighboring chains with $\varphi_m(x)=0$.
We arrive at the effective 1D Hamiltonian
\begin{eqnarray}
\nonumber
{\cal H} &=& 
\frac{\hbar v_F}{4\pi} \int dy \left(
\frac{\partial \varphi(y)}{\partial y} \right)^2
+ w \int dy \left[1-\cos{\varphi(y)}\right]\\
&-&\sum_i V_i \left[1-\cos{\left(Q y_i+\varphi(y_i) \right)}\right]
\label{eq:H-cl}
,
\end{eqnarray}
Metastable states due to
the competition between
the commensurate potential and the pinning energy
were first discussed by Abe\cite{Abe} in a different approach.
Solitons and the transition to a 3D density wave glass were
discussed by Fukuyama\cite{Fukuyama}, and bisolitons
were discussed by Larkin\cite{Larkin3} and
Ovchinikov\cite{Ov}.

\subsection{No impurity: $2\pi$-solitons}

Without impurities there exist solutions minimizing the
energy (\ref{eq:H-cl}) in which the
the phase winds by $\pm 2 \pi$ within a length
$\xi$\cite{Fukuyama,Larkin3}:
\begin{equation}
\label{eq:soli-def}
\tan{\left(\frac{\varphi(y)}{4}\right)}
=\tan{\left(\frac{\psi}{4}\right)}
\exp{\left(\pm {y\over \xi} \right)}
,
\end{equation}
where the soliton is centered at
$x_0=\pm \xi \ln{(\tan{(\psi/4)})}$.
The width of the soliton is
\begin{equation}
\label{eq:xi-def}
\xi=\sqrt{\frac{\hbar v_F}{2\pi w}}
.
\end{equation}
The energy $E_S$ of the soliton
defined by Eq.~(\ref{eq:soli-def})
is equal to $4 w\xi$\cite{Fukuyama,Larkin3}:
\begin{equation}
E_S=2 \sqrt{\frac{2 \hbar v_F w}{\pi}}
.
\end{equation}
The energy $E_S$ can be viewed as the phase ordering 
temperature. Nevertheless thermodynamic equilibrium is
not reached since the phase is frozen at a temperature
larger than $E_S$ due to collective
pinning\cite{Braso,Hosseini,Bechgaard}. At the
very low temperature considered here the relevant
excitations are local deformations of the CDW in the
form of bisolitons.

\subsection{One impurity: bisolitons}

Minimizing the
energy with respect to $\varphi(y)$ leads to the exact
expression of the phase
profile of the bisoliton associated to an impurity located
at position $y_1$ with a pinning potential $V_1$:
\begin{equation}
\label{eq:soli-1imp}
\tan{\left(\frac{\varphi(y)}{4}\right)}=
\tan{\left(\frac{\psi}{4}\right)}
\exp{\left(-\frac{|y-y_1|}{\xi}\right)}
,
\end{equation}
where $\psi$ is such that
\begin{equation}
\label{eq:match}
\sin{\left(\frac{\psi}{2}\right)}
=\frac{\pi V_1 \xi}{2 \hbar v_F} \sin{\left(\alpha_1+\psi
\right)}
,
\end{equation}
with $\alpha_1=Q y_1$. The bisoliton has a 
decay length given by Eq.~(\ref{eq:xi-def}).
There can be several solutions to the matching equation (\ref{eq:match}),
corresponding to metastable states separated by
energy barriers\cite{Larkin3,Ov}.

\subsection{A finite concentration of impurities}
\label{sec:land}

\begin{figure}
\includegraphics [width=.9 \linewidth]{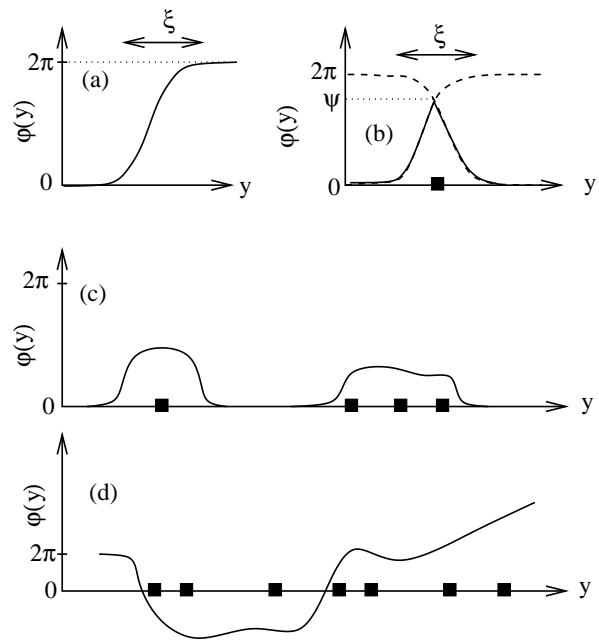}
\caption{Schematic representation of the spatial variation
of the phase $\varphi(y)$ along a chain. 
(a) A $2 \pi$ soliton varying over a length $\xi$.
(b) A bisoliton~\cite{Larkin3} (superposition of a $2\pi$-soliton
and a $2 \pi$-anti soliton) varying over a length $\xi$
generated by one strong pinning impurity. (c) Randomly
distributed bisolitons
generated by the clustering of strong pinning impurities
for $x\xi \alt 1$.
(d) Density wave glass for $x\xi \agt 1$.
The filled squares represent the impurities.
\label{fig:soli}
}
\end{figure}

\subsubsection{Transition to a 3D density wave glass}

The problem of a finite concentration of impurities was treated
in Ref.~\onlinecite{Melin-CDW} within dynamical renormalization group
(RG) describing a quench
from high temperature. The first step was to consider two bisolitons
at distance $R$. If $R$ is much larger than $\xi$ then the two
bisolitons are independent from each other and have independent
dynamics. If $R$ is much smaller than $\xi$ there is a single
bisoliton pinned by two impurities, with a longer relaxation time
than two independent solitons. The case of intermediate values of
$R$ was treated approximatively by an exponential interpolation
between the limiting cases $R \ll \xi$ and $R \gg \xi$.
In the following we consider a simplified model where the pinning
energy is additive for all values of $R$ smaller than $\xi$, 
whereas the two bisolitons are independent from each other
for all values of $R$ larger than $\xi$. Considering not only
two impurities but a finite concentration of impurities 
distributed at random in 1D, this defines clusters
of impurities: two neighboring impurities at distance $R$
belong to the same cluster if $R<\xi$, and belong to two
different clusters if $R>\xi$.

The probability to find $N_{\rm imp}\ge 1$ impurities in a given cluster
follows the exponential distribution
\begin{equation}
{\cal P}(N_{\rm imp})=\exp{(-x \xi)} \left[1-\exp{(-x \xi)}
\right]^{N_{\rm imp}-1}
,
\end{equation}
where $x$ is the impurity concentration.
The average
number of impurities within a given cluster
is equal to $\exp{(x \xi)}$. For $x \xi \agt 1$ the number of
impurities in a given cluster becomes very large so that
the phase is effectively frozen because of large energy barriers.
In this regime the system becomes a
3D density wave glass\cite{Fukuyama} and our treatment
(valid if $x\xi \alt 1$) based on the
crude 1D mean field model breaks down.

We suppose in the following that the only 3D effect is this
Larkin-Ovchinikov\cite{Larkin3,Ov}
level splitting mechanism. We start from the
limit of very dilute impurities and increase progressively the
concentration. 
The crude mean field model is expected to describe well localized
excitations in the dilute limit $x \xi \alt 1$
but at large scale and temperatures higher than the
one considered here the system
becomes a 3D elastic medium
with different properties that we do not discuss in the following. 
For instance
the impurity perturbation decays as a power-law if $D>1$
(see Ref.~\onlinecite{revue}).

\subsubsection{Multi-impurity energy landscape}

We suppose that the phase is almost constant
and equal to $\psi$ in the middle of
a bisoliton and note
$X \equiv \tan{(\psi/4)}$. 
The elastic energy corresponding to the second term of Eq.~(\ref{eq:H-cl})
is independent on
the number $N_{\rm imp}$ of impurities because it arises from the two sides
of the bisoliton profile
where $\partial \varphi(y) / \partial y$ is important.
The pinning
energy is additive because each impurity brings its own pinning energy.

The energy landscape of a cluster of $N_{\rm imp}$
impurities generalizing Ref.~\onlinecite{Melin-CDW}
is given by
\begin{eqnarray}
\label{eq:land}
&&E(X) =16 w \xi \frac{X^2}{1+X^2}\\\nonumber
&&-2 \sum_{i=1}^{N_{\rm imp}} V_i \left[
\sin{\left(\frac{\alpha_i}{2} \right)} \frac{1-X^2}{1+X^2}
+\cos{\left(\frac{\alpha_i}{2} \right)} \frac{2X}{1+X^2}
\right]^2
,
\end{eqnarray}
where the elastic and commensurate energies in the first term
are much smaller in magnitude than the pinning energy,
but play a role in lifting the degeneracy of
the effective two level system.

To illustrate Eq.~(\ref{eq:land}) we show that minimizing with
respect to $X$ for impurities at a distance much smaller than $\xi$
is equivalent to solving the sine-Gordon equation in the presence
of the pinning term. Let us consider $N_{\rm imp}$ impurities
at positions $y_1$, ..., $y_{N_{\rm imp}}$ and denote by
$\psi_1$, ..., $\psi_{N_{\rm imp}}$ the value of the phase
at the points $y_1$, ..., $y_{N_{\rm imp}}$. The solution with
the appropriate boundary conditions is
\begin{equation}
\tan{\left(\frac{\varphi(y)}{4}\right)}=
\tan{\left(\frac{\psi_1}{4}\right)} 
\exp{\left(\frac{y-y_1}{\xi}\right)}
\end{equation}
for $y<y_1$, and
\begin{equation}
\tan{\left(\frac{\varphi(y)}{4}\right)}=
\tan{\left(\frac{\psi_1}{4}\right)} 
\exp{\left(-\frac{y-y_{N_{\rm imp}}}{\xi}\right)}
\end{equation}
for $y>y_{N_{\rm imp}}$.
The derivative of $\varphi(y)$ is discontinuous at the position
of the impurities:
\begin{equation}
\frac{\partial \varphi}{\partial y}(y_k^+)
-\frac{\partial \varphi}{\partial y}(y_k^-)
=\frac{2\pi V_k}{\hbar v_F} \sin{(\alpha_k+\psi_k)}
.
\end{equation}
Assuming that $\psi$ is almost constant we obtain
\begin{equation}
\label{eq:solu-psi}
\sin{\left(\frac{\psi}{2}\right)}
=\frac{\pi\xi}{\hbar v_F} \sum_{k=1}^{N_{\rm imp}}
V_k \sin{(\alpha_k+\psi)}
,
\end{equation}
generalizing Eq.~(\ref{eq:match}).
Solving $\partial E (X)/\partial X=0$ with $E(X)$
given by Eq.~(\ref{eq:land}) and $X=\tan{(\psi/4)}$
leads directly to Eq.~(\ref{eq:solu-psi}).

\subsection{Properties of the energy landscape}
\begin{figure}
\includegraphics [width=.9 \linewidth]{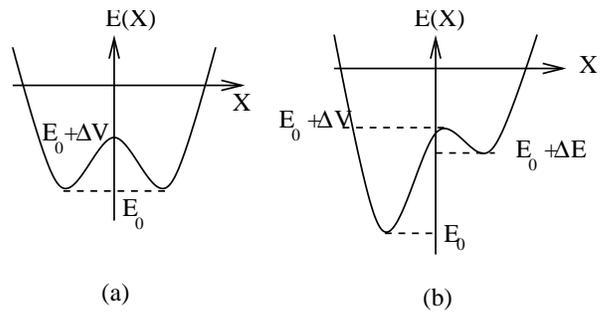}
\caption{Schematic representation of the energy landscape
in the commensurate case (a) and in the incommensurate case (b).
The ground state is at energy $E_0$. The metastable state
is at energy $E_0+\Delta E$, where $\Delta E$ is the splitting.
The unstable ``bounce'' state is at energy $E_0+\Delta V$
where $\Delta V$ is the energy barrier.
\label{fig:land}
}
\end{figure}

The energy landscape $E(X)$ given by (\ref{eq:land})
describes a ground state, separated from a metastable state
by an energy barrier. There might be
more than two energy minima for some realizations of disorder\cite{Larkin3,Ov}.
In this
case we restrict the energy landscape to the ground state and to
the energy minimum separated from the ground state by the lowest
energy barrier.
We note $\Delta E$ the difference between
the energies of the metastable state and the ground state
and $\Delta V$ the energy barrier (see Fig.~\ref{fig:land}).

\subsubsection{Commensurate case}

\begin{figure}
\includegraphics [width=.9 \linewidth]{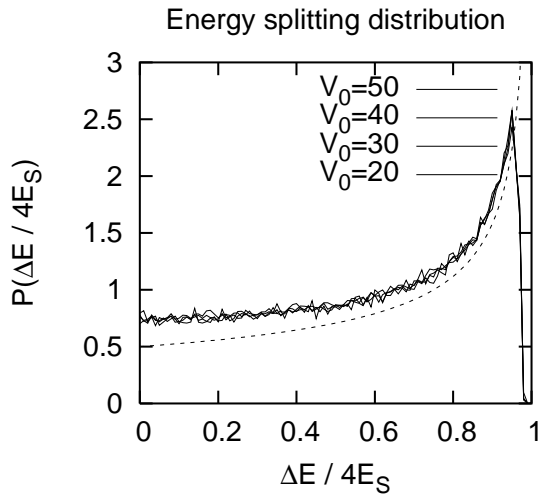}
\caption{Distribution of level splitting $\Delta E$
for the model with random phases discussed in
section~\ref{sec:land}. The impurities
are distributed at random on a chain of length $\xi$.
The impurity concentration is such that
the average number of impurities is
$\overline{N}_{\rm imp}=x \xi=1$. The curves corresponding to
$x \xi=0.5$ and $x \xi=5$ are superposed on the curves
corresponding to $x \xi=1$.
The pinning potentials $V_i$ are uniformly distributed
in the interval $\left[0,V_0\right]$ with $V_0$ indicated
on the figure. The distribution of level splitting
normalized to $4 E_S$ is almost independent on
$w \xi$ and $V_0$, chosen such that
$4 E_S \ll V_0$. The dashed line corresponds
to $P(z)=1/[2\sqrt{1-z}]$, with $z=\Delta E / 4 E_S$
(see Eq.~\ref{eq:PE}).
\label{fig:dis-spl}
}
\end{figure}

In the dimerized case ($Q=\pi$, $\alpha=0$) $\Delta E$
is equal to zero since the energy landscape is
symmetric under  a change of sign of the ``coordinate'' $X$.
This can be seen from Eq.~(\ref{eq:land})
by noting that $\cos{(\alpha_i/2)}=\pm 1$,
$\sin{(\alpha_i/2)}=0$ if $Q x_i=2\pi n$, with $n$ an integer,
and $\cos{(\alpha_i/2)}=0$,
$\sin{(\alpha_i/2)}=\pm 1$ if $Q x_i=(2 n+1)\pi$,
with $n$ an integer. 
The degeneracy can also be seen directly from Eq.~(\ref{eq:solu-psi})
that, in the commensurate case $Q=\pi$, becomes
\begin{equation}
\cos{\left(\frac{\psi}{2}\right)}=
\left[\frac{2 \pi \xi}{\hbar v_F}
\sum_{k=1}^{N_{\rm imp}}
V_k \cos{\alpha_k}\right]^{-1}
.
\end{equation}
If $\psi$ is a solution then $-\psi$ is also a solution,
explaining the degeneracy. The degeneracy can be also understood
by noting that the transformation
$\varphi(x) \rightarrow -\varphi(x)$ is a symmetry of the
pure system, that is preserved by the pinning term for
commensurate impurities but
not for incommensurate impurities.

In the commensurate case
it is thus not possible
to communicate energy 
over long time scales 
to the effective two-level system
by increasing temperature.
The classical model predicts no slow relaxation at
all in the commensurate case whereas in experiments there exists
slow relaxation, even though faster than in the incommensurate
compound\cite{new-ref}. 
Adding quantum tunneling between the two energy minima of the energy landscape
can generate two non degenerate energy levels corresponding to 
symmetric and antisymmetric wave functions, therefore restoring a finite
heat response. 

\subsubsection{Energy splitting distribution in the incommensurate case}

The distribution of energy splitting is shown on
Fig.~\ref{fig:dis-spl}.
The distribution of splitting is close to
\begin{equation}
\label{eq:PE}
P(\frac{\Delta E}{4 E_S}) \simeq
\frac{1}{2} \left(1-\frac{\Delta E}{4 E_S}\right)^{-1/2}
.
\end{equation}
The most probable level spacing is
close to $4 E_S$, independent on the value of the pinning
potential and
on the number of impurities involved in the bisoliton, showing that
the shape of $P(\Delta E/4 E_S)$ is almost unchanged
when the concentration of impurities increases. 
The energy splitting distribution (\ref{eq:PE}) can be understood
in the case of a single soliton by noting that the pinning energy
is much larger than the elastic energy so that the elastic term
can be treated as a perturbation. There are four values of $X$
minimizing the pinning energy:
\begin{eqnarray}
\label{eq:X_0}
X_0^{(\epsilon)}&=&\frac{\epsilon-\sin{(\alpha/2)}}{\cos{(\alpha/2)}}\\
X_1^{(\epsilon)}&=&\frac{\epsilon+\cos{(\alpha/2)}}{\sin{(\alpha/2)}}
\label{eq:X_1}
,
\end{eqnarray}
with $\epsilon=\pm 1$. The pinning energy of the solution
(\ref{eq:X_0}) is $E_{\rm pin}(X_0^{(\epsilon)})=-2 V$,
and the pinning energy of the
solution (\ref{eq:X_1}) is $E_{\rm pin}(X_1^{(\epsilon)})=0$.
The elastic energy of the solution $X_0^{(\epsilon)}$
is $E_{\rm el}(X_0^{(\epsilon)})=2 E_S (1-\epsilon \sin{(\alpha/2)})$.
The energy splitting $E_{\rm el}(X_0^{(-)})-E_{\rm el}(X_0^{(+)})$
is then distributed according to Eq.~(\ref{eq:PE}) since
$\alpha$ is uniformly distributed.

The existence of the upper bound $4 E_S$
in the energy splitting distribution
is compatible with the experimental observation of
a high temperature
tail $C_v \sim 1/T^2$ of a Schottky anomaly in the equilibrium
specific heat, with a well-defined level splitting.
Experimentally the level splitting $\Delta E_0$ of a two-level
system is related to the temperature $T_{\rm max}$ of the maximum of the
Schottky anomaly by the relation $\Delta E_0 \simeq 2.5 k_B T_{\rm max}$.
It was shown experimentally that
$T_{\rm max}<30$~mK\cite{manips-PF6-JPCM-I} so that
$\Delta E_0=4 E_S \alt 100$~mK. The existence of a well-defined
level splitting in experiments is a universal property, valid
for commensurate and incommensurate systems, and independent
on the value of the CDW or SDW critical temperature that can
vary by more than one order of magnitude from one compound to
the other ($T_{\rm Peierls}=218$~K for o-TaS$_3$,
$T_{\rm SDW}=12$~K for (TMTSF)$_2$PF$_6$,
$T_{\rm SP}=15$~K for the spin-Peierls compound (TMTTF)$_2$PF$_6$,
$T_{\rm AF}=13$~K for the antiferromagnet (TMTTF)$_2$Br).
This is compatible with the fact that $\Delta E_0$ is related only
to the strength $w$ of interchain interactions and the Fermi velocity $v_F$:
\begin{equation}
\Delta E_0 = 4 E_S = \frac{16}{\sqrt{2 \pi}} \sqrt{\hbar v_F w}
,
\end{equation}
where $v_F$ has the same magnitude for several compounds
(for instance $v_F=0.86 \times 10^7$cm.sec$^{-1}$ in
(TMTSF)$_2$PF$_6$\cite{Gruner}) and interchain couplings are also expected
to take similar orders of magnitude, from what we deduce that $\Delta E_0$
is almost identical in all samples, independent on the charge or spin gap.
The energy $\Delta E_0$ is proportional to the elastic energy $E_S$ of a
$2 \pi$-soliton in the absence of impurities, not to be confused with the
activation energy $E_A$ of the order of $100$~K probed in transport
experiments\cite{Takoshima}.
There is thus no inconsistency in the difference between the
orders of magnitude of $E_A$ and $E_S$.

\subsubsection{Distribution of energy barriers}

\begin{figure}
\includegraphics [width=.9 \linewidth]{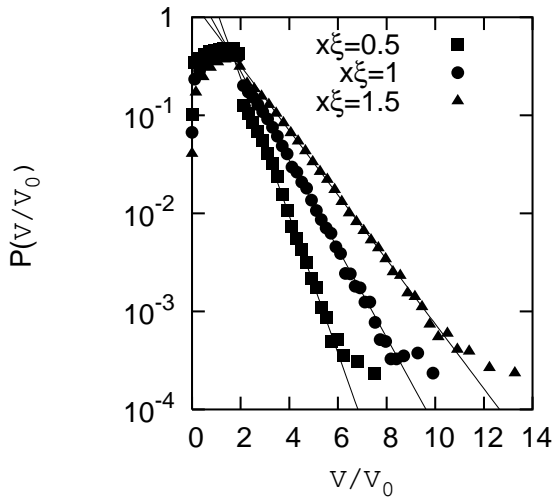}
\caption{Distribution of the dimensionless energy barrier
$\Delta V/V_0$ in the cluster model. We use
$x \xi = 0.5,1,1.5$, $k_F=\pi/2+0.02$.
The pinning potentials $V_i$ are uniformly distributed
in the interval $\left[0,V_0\right]$ with $V_0=50$.
The solid lines correspond to the fits
$P(\Delta V/V_0)=1.45\times
\exp{(-0.76\times\Delta V/V_0)}$ for $x\xi=0.5$,
$P(\Delta V/V_0)= 2.25 \times \exp{(-1.04 
\times\Delta V/V_0)}$ for $x\xi=1$, and
$P(\Delta V/V_0)= 6 \times \exp{(-1.61 \times
\Delta V/V_0)}$ for $x\xi=1.5$.
\label{fig:dis-bar}
}
\end{figure}

The distribution
of energy barriers $P(\Delta V/V_0)$ is well fitted
by an exponential function (see Fig.~\ref{fig:dis-bar}).
The average relaxation time defined as
\begin{equation}
\tau_0 \int d (\Delta V)
P(\Delta V) \tau_0 \exp{(\Delta V/T)}
,
\end{equation}
with $T$ the temperature
diverges at a finite temperature, like in the REM-like
trap model~\cite{trap-model}. However this model,
useful for discussing waiting time effects, 
over evaluates glassiness compared to the dynamical RG
already discussed in Ref.~\onlinecite{Melin-CDW}
(see section~\ref{sec:spec-rel}).

\subsection{Out-of-equilibrium specific heat}

\begin{figure}
\includegraphics [width=.9 \linewidth]{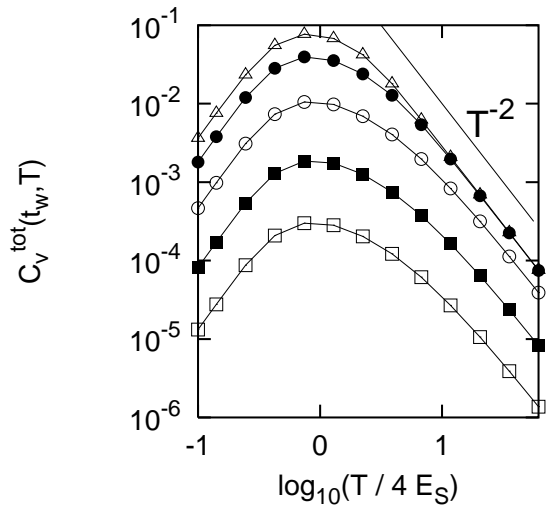}
\caption{Variation of the out-of-equilibrium total specific heat
as a function of $\log_{10}(T/4 E_S)$, for
different values of $t_w/\tau_0$:
$\log_{10}(t_w/\tau_0)=-2$ ($\square$),
$\log_{10}(t_w/\tau_0)=-1.2$ ($\blacksquare$),
$\log_{10}(t_w/\tau_0)=-0.4$ ($\circ$),
$\log_{10}(t_w/\tau_0)=0.4$ ($\bullet$).
$\log_{10}(t_w/\tau_0)=1.2$ ($\triangle$).
We used $x \xi=1$.
The pinning potentials $V_i$
are uniformly distributed in the interval $[0,V_0]$ with
$V_0/E_S=12.5$. The specific heat for long $t_w/\tau_0$
follows a $1/T^2$ behavior for $T/4 E_S \agt 1$,
where $E_S$ is the energy of a soliton.
The solid line represents the $C_v \sim T^{-2}$ behavior.
\label{fig:Cv}
}
\end{figure}

To evaluate energy relaxation in the incommensurate case,
we restrict the energy landscape to the ground state, the
metastable state and the unstable ``bounce'' state at the top of
the barrier. We are left with a
two-state trap model with energies
$-E_1$ and $-E_2$ (with $E_1>E_2$).
$E_1$ is equal to the barrier $\Delta V$ calculated in the preceding section
and $E_2$ is equal to the barrier $\Delta V$ minus the splitting $\Delta E$. 
The unstable state is at zero energy.
We note $P_{1}(t)$ ($P_2(t)$) the probability to be in state ``1''
(``2'') at time $t$ and note $\tilde{P}_{1,2}(t)=P_{1,2}(t)
\exp{(-E_{1,2}/2T)}$. The evolution of the probabilities is given by
Glauber dynamics\cite{Glauber} 
\begin{equation}
\tau_0\frac{d}{dt}  \left[ \begin{array}{c}
\tilde{P}_1 \\ \tilde{P}_2 \end{array} \right]= \hat{G}
\left[ \begin{array}{c}
\tilde{P}_1 \\ \tilde{P}_2 \end{array} \right]
,
\end{equation}
with
$G_{1,1}=-\exp{(-E_1/T)}$, $G_{2,2}=-\exp{(-E_2/T)}$,
$G_{1,2}=G_{2,1}=\exp{(-(E_1+E_2)/2T)}$. 
Experimentally the time $\tau_0$
associated
to bisoliton dynamics is of order of $1$~sec for the thermally activated
process.
The time dependence of the occupation probabilities $P_1(t)$
and $P_2(t)$ is obtained
by diagonalizing the 2$\times$2 matrix $\hat{G}$, from what
we deduce the value of the energy $U(t_w,t_w+\tau,T)$ as a function
of the waiting time $t_w$, the time $\tau$ elapsed since the
waiting time and temperature $T$.
The out-of-equilibrium total specific heat is defined as the total
heat released divided by the temperature variation:
\begin{equation}
C_v^{\rm tot}(t_w,T) = \frac{U(t_w,t_w,T+\Delta T)-U(0,0,T)}{\Delta T}
.
\end{equation}
The dynamics equations can be solved exactly
in the case of a single two-level system with energies $-E_1$
and $-E_2$ and expanded in the high temperature 
regime $T \gg |E_1-E_2|$, leading
to the specific heat
\begin{eqnarray}
C_v&=&\frac{(E_1-E_2)^2}{4 T^2 } \\\nonumber
&\times& \left[1-\exp{\left[- \frac{t_w}{\tau_0}
\left[\exp{\left(-\frac{E_1}{T}\right)}
+\exp{\left(-\frac{E_2}{T}\right)}\right]\right]}
\right]
,
\end{eqnarray}
proportional to $1/T^2$, as expected for the
high temperature tail of a Schottky
anomaly. Increasing the waiting time increases the energy
transfered to the two-level system and therefore increases the
amplitude of the $1/T^2$ term.

The variations of the out-of-equilibrium total specific heat
$C_v^{\rm tot}(t_w,T)$ as a function of temperature $T$
for different values of the waiting time $t_w$ are shown
on Fig.~\ref{fig:Cv} for a finite concentration of
incommensurate impurities. 
At temperatures larger than the maximal splitting
$\Delta E=4 E_S$ 
the out-of-equilibrium specific heat follows a $1/T^2$
behavior.
The out-of-equilibrium specific heat is
strongly reduced as the waiting time decreases but still
follows a $1/T^2$ behavior, in
a qualitative agreement with
experiments
(see Fig. 3 in Ref.~\onlinecite{manips-PF6-JPCM-I}).

\subsection{Spectrum of relaxation times}
\label{sec:spec-rel}
\begin{figure}
\includegraphics [width=.9 \linewidth]{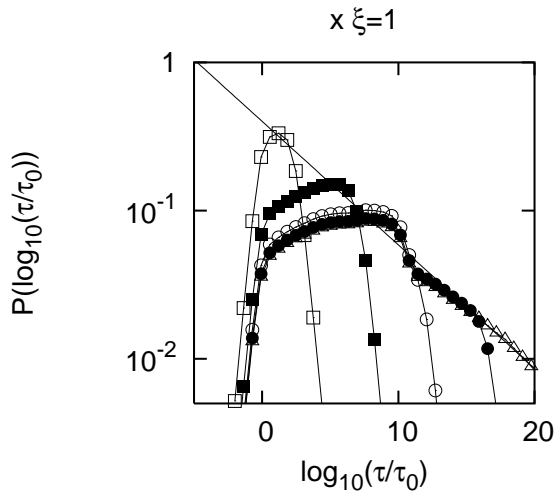}
\caption{Spectrum of relaxation times deduced from 
energy relaxation (see Eq.(\ref{eq:spectrum})).
We use $V_0/4 E_S=12.5$, $T/4 E_S=400$,,
and $x \xi=1$.
The pinning potentials $V_i$
are uniformly distributed in the interval $[0,V_0]$.
The waiting times correspond to
$\log_{10}(t_w/\tau_0)=2.4$ ($\square$), $7.2$ ($\blacksquare$),
$12$ ($\bullet$), $16.8$ ($\circ$), $21.6$ ($\triangle$).
The solid line is a fit to $\log_{10}(P(\log_{10}{\tau}))
=a+b \log_{10}{\tau}$, with $b=-0.083$.
\label{fig:sp}
}
\end{figure}

\begin{figure}
\includegraphics [width=.9 \linewidth]{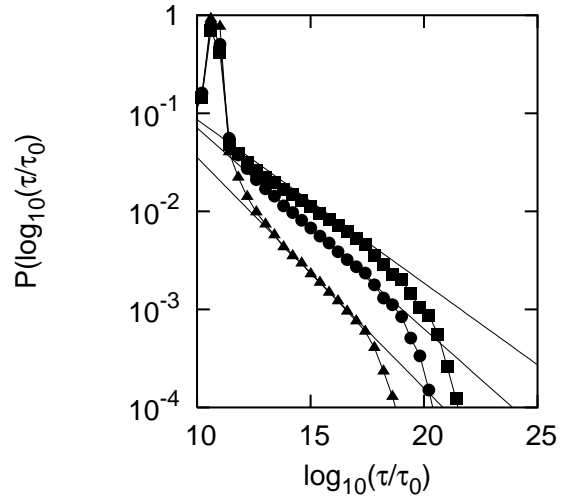}
\caption{Spectrum of relaxation times deduced from
dynamical renormalization group.
We use $V_0/4 E_S=12.5$, $T/4 E_S=400$,,
and $x \xi=1.5$ ($\blacksquare$),
$x\xi=1$ ($\bullet$) and $x\xi=0.5$ ($\blacktriangle$).
The pinning potentials $V_i$
are uniformly distributed in the interval $[0,V_0]$.
The solid line is a fit to $\log_{10}(P(\log_{10}{\tau}))
=a+b \log_{10}{\tau}$, with $b=-0.23$ for $x\xi=1.5$,
$b=-0.21$ for $x\xi=1$ and $b=-0.17$ for $x\xi=0.5$.
We carried out simulations with larger system sizes
and found practically no finite size effects on
the data on this figure.
\label{fig:spec2}
}
\end{figure}

Let $U(t_w,t_w+\tau,T)$ be the energy of the two-level
system at time $t=t_w+\tau$, with a heat pulse applied between
$t=0$ and
$t=t_w$. The spectrum of relaxation times is deduced from
$U(t_w,t_w+\tau,T)$ by assuming that
\begin{equation}
U(t_w,t_w+\tau,T)=\int P_{t_w}(\ln{\tau'}) \exp{(-\tau/\tau')}
d \ln{\tau'}
.
\end{equation}
Replacing the exponential by a step-function leads to
the relaxation time spectrum\cite{spin-glass1}
\begin{equation}
\label{eq:spectrum}
P_{t_w}(\ln{\tau}) \propto \frac{\partial U(t_w,t_w+\tau,T)}
{\partial \ln{\tau}}
.
\end{equation}
We have shown on
Fig.~\ref{fig:sp} the spectra of relaxation times
$P_{t_w}(\ln{\tau})$ for $x \xi=1$. Similar results are obtained
for $x\xi=0.5$ and $x\xi=1.5$, with however a different value of
the exponent of the power-law.
The long time tail of the spectrum is a power-law, in agreement
with experiments (see Ref.~\onlinecite{Melin-CDW}). As the waiting
time increases the power-law regime extends to ever longer times
without limits, compatible with the existence of a genuine glass
transition that is an artifact of this clustering model.

For comparison we have also calculated the relaxation
time spectra obtained from dynamical RG (see Fig.~\ref{fig:spec2}).
The calculation is identical to Ref.~\onlinecite{Melin-CDW}:
we suppose a quench 
from high temperature at time $t=0$. The small time degrees of freedom
are progressively eliminated, making bigger object that relax
more slowly. In this approach the energy landscape of two impurities
at distance $R$ is interpolated between the two limiting cases
$R \ll \xi$ and $R \gg \xi$ so that the phase fluctuations are
more important compared to the previous clustering model.
Impurities can thus be depinned in sequence, not necessarily
all together.
We obtain also a power-law relaxation but 
(i) the exponent of the power-law spectrum of relation times is different
from the clustering model (smaller relaxation times are favored), and
(ii) ageing is ``interrupted'' for the clustering model (there exists
a maximal relaxation time). As expected the clustering model discussed
previously over evaluates glassiness but is nevertheless useful for addressing
qualitatively waiting time effects that could not be discussed within
dynamical RG.
Experiments favor interrupted ageing
with respect to a genuine dynamical glass transition (see
Ref.~\onlinecite{Melin-CDW} for an analysis of experiments based on
a REM-like model that is justified microscopically by the 
dynamical RG approach).

\section{Quantum limit (substitutional and strong pinning impurities)}
\label{sec:quantum}

\begin{figure}
\includegraphics [width=.9 \linewidth]{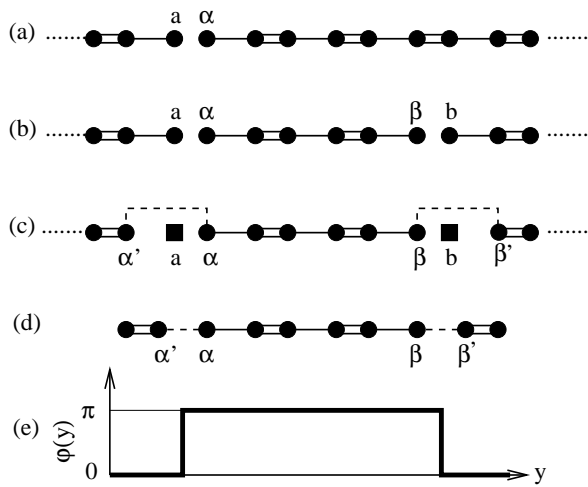}
\caption{Schematic representation of a dimerized CDW chains
with open boundary conditions with one impurity (a) or
two impurities (b). The double lines represent the strong
bonds $t+\epsilon$.
(b) corresponds to $R=y_\beta-y_\alpha$ odd. We represented on (c)
two substitutional impurities (black square) leaving
unpaired fermions at sites $\alpha$ and $\beta$.
(d) corresponds to the chain where the impurity sites
have been removed\cite{FM}.
(e) shows the spatial variation of the phase
corresponding to the chain (d).
\label{fig:dim}
}
\end{figure}

\subsection{Preliminaries}

We base the discussion in this section on the electronic part of the
Peierls Hamiltonian. We discuss a model of disordered CDW 
there is only a charge sector, or a model of disordered
SDW where there is only a spin sector. We expect this simplified
models to capture part of the properties of real compound in which
there is a gap in both the spin and charge channels.
The analogy between the CDW case with only a charge channel and
the SDW case with only a spin channel is
based on the fact that one dimensional tight binding models can
be mapped onto XX magnets in one dimension through a Jordan-Wigner
transformation. It was shown explicitely in the dimerized case
that the properties of substitutional disorder are almost identical for
the XX and Heisenberg models\cite{FM} in spite of interactions between 
Jordan-Wigner fermions in the Heisenberg case. We expect this similarity to
be valid also in the incommensurate case. 

Quantum effect were already discussed in Ref.~\onlinecite{Braso}
but here we work within a different microscopic model that was proved to
be useful in the case of doped spin-Peierls systems\cite{FM},
and generalize it to the case of incommensurate CDWs.
Substitutional disorder in a spin-Peierls system
(relevant to
Cu$_{1-x}$Zn$_x$GeO$_3$\cite{CuGeO3-1,CuGeO3-2,CuGeO3-3,CuGeO3-4})
is qualitatively equivalent to
impurities that break the chains into finite length
segments (see Ref.~\onlinecite{FM} and  section~\ref{sec:dim}).
We have shown on Fig.~\ref{fig:dim} a schematic representation of
substitutional impurities in dimerized chains. 
The phase of the dimerization changes by $\pi$ in the chain
where the site of the impurity has been removed. For a CDW or SDW with
a wave-vector $Q=2 k_F$ the phase
changes by $Q a_0$ once the impurity sites have been removed, where $a_0$ is
the lattice parameter. The absolute value of the amplitude of the CDW 
is supposed to
remain constant. The solitons generated by the substitutional disorder that
we consider here are thus {\it phase solitons},
connecting two degenerate ground states with different phases.
One can cross-over continously from edge states in the
limits $t_{\alpha,\alpha'}=t_{\beta,\beta'}=0$ to solitons connecting
two ground states as the parameters $t_{\alpha,\alpha'}$
and $t_{\beta,\beta'}$ are increased.

There is a flat classical energy landscape
in the case of open chains for the
model defined by Eq.(\ref{eq:H-cl}) in the limit $w=0$.
The classical Hamiltonian is just (\ref{eq:H-cl}) on an open chain,
without the pinning term:
\begin{eqnarray}
{\cal H} &=&
\frac{\hbar v_F}{4\pi} \int_0^L dy \left(
\frac{\partial \varphi(y)}{\partial y} \right)^2\\
&+& w \int_0^L dy \left[1-\cos{\varphi(y)}\right]
\nonumber
,
\end{eqnarray}
that has no bisoliton metastable states like in the
preceding section (the ground state is $\varphi(y)=0$).
The bound state associated to a single substitutional impurity
in the quantum model defined by the Peierls Hamiltonian
\begin{equation}
\label{eq:H-elec}
\label{eq:H-Peierls}
{\cal H} = \sum_i
\left[ t + \epsilon \cos{(2 k_F y_i)} \right]
\left[ c_{i+1}^+ c_i + c_i^+ c_{i+1} \right]
,
\end{equation}
is thus at the lowest
possible energy (exactly in the middle of the gap).
The variable $t$ in Eq.~(\ref{eq:H-Peierls})
is equal to the average hopping amplitude and $\epsilon$ 
is the amplitude of the modulation.
The variable $y_i$ in
Eq.(\ref{eq:H-elec}) is the coordinate of the site
number $i$: $y_i=i a_0$, with $a_0$ the lattice parameter.
The bound state due to substitutional disorder can
be occupied by an electron and a hole with an equal 
probability, and there is a fast dynamics of the
occupation probability since there are no energy barriers
in the classical limit. Like in spin glasses mostly
governed by RKKY interactions it is expected that
the ``microscopic time'' $\tau_0$ is extremely small
($\tau_0\simeq 10^{-12}$~sec in spin glasses). 
By contrast there is slow dynamics with $\tau_0 \simeq 1$~sec
in the case of the organic spin-Peierls compound (TMTTF)$_2$PF$_6$,
well described by the classical model discussed previously.

The width of a soliton in the quantum limit is equal to 
\begin{equation}
\xi_0(E)=\frac{2 \hbar v_F}{\sqrt{\Delta^2-E^2}}
,
\end{equation}
where $E$ is the energy,
$\Delta$ is equal to the Peierls gap and
$v_F$ is the Fermi velocity in the absence of lattice
distortions.
$\xi_0$ is
different from the soliton or bisoliton width in the classical limit
$\xi=\sqrt{\hbar v_F/2 \pi w}$
discussed in the preceding section.
In the classical model one should incorporate interchain interactions
for the soliton or bisoliton to have a finite width whereas
in the quantum model
the zero energy soliton width is finite in the absence of
interchain interactions.
Experimentally the slow relaxation properties associated to the
$1/T^2$ tail of the specific heat are independent on the value
of the gap $\Delta$ and of the transition
temperature that can vary by more than one order of
magnitude.
In the case of the SDW (TMTSF)$_2$PF$_6$ the BCS-like relation
$\Delta \simeq 1.7 T_c$ is well verified ($T_c=11.5 \div 12$~K
and $\Delta=20$~K from NMR \cite{Nad}). The Fermi velocity\cite{Gruner}
is $v_F \simeq 0.86 \times 10^7$cm.sec$^{-1}$
so that $\xi_0=2 \hbar v_F/\Delta \simeq 90 a_0$, with the lattice parameter
$a_0=7.3 \AA$. 
In the case of o-TaS$_3$ we do not have a precise data for $v_F$ and we
take $v_F = 10^7$cm.sec$^{-1}$ as a typical value. We have 
$T_c^{({\rm CDW})}=215$K and $\Delta_{\rm CDW} =3.73 T_c \simeq 780$K so that
$\xi_0^{({\rm CDW})}=6 a_0$
with $a_0=3.34 \AA$. However it is likely that there is also a smaller
spin gap since the spin susceptibility follows a power-law.
An energy scale $E=30$~K can be estimated from the deviations of the
power-law. Assuming that the power-law susceptibility is due to randomly
distributed magnetic moments we identify the energy scale $E$ to the
average exchange energy $J_{\rm av}=\Delta x \xi$ with $x \xi=0.3$ estimated
from the power-law susceptibility (see section~\ref{sec:power}) so
that $\xi_0^{({\rm spin})}=150 \div 450 a_0$.
$\xi_0^{({\rm spin})}$ in the SDW compound (TMTSF)$_2$PF$_6$ and o-TaS$_3$
are thus one order of magnitude
larger than $\xi_0$ in the spin-Peierls compound CuGeO$_3$.

\subsection{Strong pinning impurities}
\subsubsection{Dyson matrix}
The quantum model given by Eq.~(\ref{eq:H-Peierls}) can also be
used to treat strong pinning impurities, not only substitutional
disorder.
The Hamiltonian is equal to (\ref{eq:H-Peierls}), plus a
term describing the impurity potential:
\begin{equation}
H_{\rm imp} = - \sum_i V_{y_i} c_{y_i}^+ c_{y_i}
,
\end{equation}
where the strong pinning impurities are located at
random positions $\{y_1,...,y_{\rm N_{\rm imp}}\}$. We suppose that $V_{y_i}=V$
is the same for all impurities and that $V>0$.
We note $G(E)$ the Green's function at energy $E$
in the presence of the
pinning potential and $g(E)$ the Green's function in the
absence of pinning potential. The Green's function
$G(E)$ is obtained from inverting the Dyson matrix:
\begin{equation}
\sum_{k=1}^{N_{\rm imp}}\left[ \delta_{i,k} +
g_{y_i,y_k}(E) V_{y_k}
\right] G_{y_k,y_j}(E) = g_{y_i,y_j}(E)
,
\end{equation}
where the Green's function
$g_{y_i,y_k}(E)$ is given in Appendix~\ref{app:Green}.
The bound state energies correspond to the poles of $G(E)$.

\subsubsection{Two pinning centers}
\begin{figure}
\includegraphics [width=.9 \linewidth]{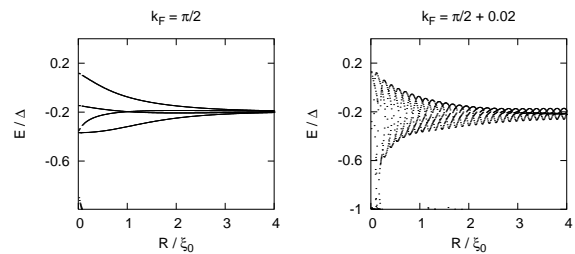}
\caption{Evolution of the bound state energy levels as
a function of the distance between the two pinning
centers, in the commensurate case $k_F=\pi/2$ (a)
and in the incommensurate case $k_F=\pi/2+\delta k_F$,
with $\delta k_F=0.02$ (b). The energy is in units of
$\Delta = \sqrt{2} \epsilon$ and the distance is in units
of $\xi_0=2 v_F / \Delta$. We used $\epsilon/t=10^{-3}$
and $V/t=10$.
\label{fig:2sol}
}
\end{figure}

In the
commensurate case $k_F a_0=\pi/2$ the energy levels evolve
smoothly as a function of the distance between the two impurities
(see Fig.~\ref{fig:2sol}-(a)). In the incommensurate case
$k_F a_0 = \pi/2 + \delta k_F$ (see Fig.~\ref{fig:2sol}-(b))
the energy levels fluctuate strongly as the distance between
the two impurities is reduced. This shows that 
quantum mechanical interactions among impurities
enhance disorder effects in the incommensurate case but not in the
commensurate case. 

\subsection{Substitutional disorder}
\label{sec:dim}
\label{sec:RG}

\subsubsection{Open chains in the dimerized limit}

Let us consider two semi-infinite dimerized chains ending at
sites ``a'' and ``$\alpha$'' (see Fig.~\ref{fig:dim}-(a)).
We note $t_0=t_{a,\alpha}$ the
value of the hopping between sites ``a'' and ``$\alpha$''
(equal to $t+\epsilon$ or $t-\epsilon$) in the infinite chain.
The Dyson equation relates the 
Green's functions $g_{i,j}$ of the infinite chain to
the Green's functions $G_{i,j}$ of the
semi-infinite chain:
\begin{eqnarray}
\label{eq:Dy1}
g_{a,a}(E) &=& G_{a,a}(E) + t_0^2 G_{a,a}(E)
g_{a,a}(E)\\
g_{\alpha,\alpha}(E) &=& G_{\alpha,\alpha}(E)
+t_0^2 G_{a,a}(E) g_{\alpha,\alpha}(E)
\label{eq:Dy2}
,
\end{eqnarray}
with $g_{a,a}(E) = g_{\alpha,\alpha}(E) =
E/(2 t \sqrt{2 \epsilon^2-E^2})$. The solution
of (\ref{eq:Dy1}) and (\ref{eq:Dy2}) is
\begin{equation}
G_{a,a}(E) = \frac{t}{t_0^2}
\frac{\sqrt{2 \epsilon^2-E^2}}{E}
\left\{-1+\eta_a \sqrt{1+\frac{t_0^2}{t^2}
\frac{E^2}{2\epsilon^2-E^2}} \right\}
,
\end{equation}
with $\eta_a=\pm 1$. In the case of two weak bonds
in the chain (see Fig.~\ref{fig:dim}-(b)) we find
$G_{\alpha,\beta}=g_{\alpha,\beta}/{\cal D}$, with
\begin{eqnarray}
{\cal D} &=& \left[1+t_{a,\alpha}^2 G_{a,a} g_{\alpha,\alpha} \right]
\left[1+t_{b,\beta}^2 G_{b,b} g_{\beta,\beta} \right]\\
&-& t_{a,\alpha}^2 t_{b,\beta}^2 G_{a,a} G_{b,b}
g_{\alpha,\beta} g_{\beta,\alpha}
\nonumber
.
\end{eqnarray}
If the separation between the
boundaries is large enough the energy of the bound states
can be expanded in the small parameter $\exp{(-R/\xi_0)}$ where
$\xi_0/a_0=2 \sqrt{2} t/\epsilon$ is the zero energy coherence
length. Since the bound states are close to the Fermi energy we
also expand $G_{\alpha,\beta}$ in powers of $E$.
If $R=y_\beta-y_\alpha$ is even we find that one bound state
is generated at one end of the chain. In the case
$R$ odd on Fig.~\ref{fig:dim}-(b) we find two bound states
at $E^{(\pm)}=\pm (\epsilon/3) \exp{(-R/\xi_0)}$
corresponding to $\eta_a=\eta_b=-1$. 
In the ground state $E^{(-)}$ is occupied with an electron
and $E^{(+)}$ is occupied with a hole.
The lowest neutral
excitation corresponds to an electron in the
level $E^{(+)}$ and a hole
in the level $E^{(-)}$. This is in agreement with a previous
model developed for the spin-Peierls compound
Cu$_{1-x}$Zn$_x$O$_3$\cite{FM} as well as in a qualitative
agreement with numerical
simulations\cite{Laukamp,Poilblanc}.

\subsubsection{Substitutional disorder in CDWs and SDWs}
\label{sec:subst-cdw}
\label{sec:power}
\begin{figure}
\includegraphics [width=.9 \linewidth]{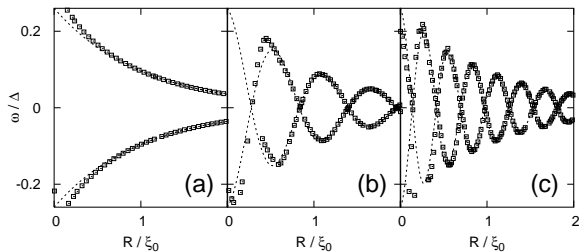}
\caption{Evolution of the bound state energies as a function of
the distance $R=|y_a-y_b|$ between the two impurities.
We use $k_F=\pi/2+\delta k_F$, with
$\delta k_F=0$ (a), $\delta k_F=0.02$ (b) and
$\delta k_F=0.04$ (c). We use the parameters
$\epsilon/t=10^{-2}$. $t_{\alpha,\alpha'}/t=t_{\beta,\beta'}/t=1/2$.
The dashed line is a fit to $E/\Delta=\pm 0.26 
\cos{((\delta k_F) R)} \exp{(-R/\xi_0)}$.
\label{fig:RKKY}
}
\end{figure}

We consider now a more realistic model of substitutional disorder, both for
commensurate and incommensurate CDWs. We suppose that two
substitutional impurities ``remove'' the sites ``a'' and ``b'' from the
chain and that second neighbor interactions $t_{\alpha,\alpha'}$
and $t_{\beta,\beta'}$ couple the right and left neighbors
(see Fig.~\ref{fig:dim}-(c)). The model is similar to
Ref.~\onlinecite{FM} proposed for Zn impurities in CuGeO$_3$
except that we consider here the
incommensurate case.
We calculate the poles
and residues as a function of the distance $R=|y_a-y_b|$
between the two impurities (see Fig.~\ref{fig:RKKY}).
We use $k_F=k_F^{(0)}+ \delta k_F$, with
$k_F^{(0)}=\pi/2$ and show on Fig.~\ref{fig:RKKY}
the Friedel oscillations of the bound state levels for odd values of $R$.
The phase of the Friedel oscillations is shifted by $\pi/2$ for even
values of $R$.
We deduce from the level crossings that there exits a 
change of sign in the hopping amplitude between the two
solitons as $R$ is increased.

Similarly to a model of doped spin-Peierls system\cite{FM}
substitutional disorder in CDWs and SDWs
provides an explanation to the power-law specific heat
$C_v \sim T^{\alpha}$ observed
in disordered CDWs and SDWs, and to the susceptibility $\chi(T)
\sim T^{-1+\alpha}$ observed in the CDW o-TaS$_3$\cite{Bilja-2003}
(with the same value of $\alpha$ in the specific heat and
susceptibility experiments).
In experiments on the SDW (TMTSF)$_2$PF$_6$\cite{manips-PF6},
on changing the time constant
in a heat pulse experiment by a factor of $100$, there is an increase
by a factor of $7$ of the amplitude of the $T^{-2}$ contribution whereas
the $T^\alpha$ contribution changes only by $20 \%$, and not in a
systematic manner. In the commensurate (TMTTF)$_2$Br compound and also in
a heat
pulse experiment, an increase of the time constant by a factor of $3$
results in an increase of the $T^{-2}$ term by a factor of $3$
and only an increase of the $T^\alpha$ term by $20 \%$. This shows that
the $T^\alpha$ contribution to the specific heat can be interpreted as
an equilibrium property, which is the case if frustration due to
Friedel oscillations can be neglected as in a dilute one dimensional
model. Focusing on the SDW case
we consider only exchanges $J(R)$ between nearest neighbor
solitons and disregard the changes of sign in $J(R)$.
The exchange distribution scales like $P(|J|) \sim |J|^{-1+x \xi_0}$,
where $x$ is the concentration of substitutional impurities\cite{FM}.
The energy of ``active'' pairs of spins
is approximately given by $U=\int_0^T J P(J) d J$, where $T$
is the temperature so that the temperature dependence of the
specific heat is given by $C_v \sim T^{x \xi_0}$.
The susceptibility is approximately given by a Curie contribution for the
fraction of ``active'' spin having an energy smaller than $T$ so
that the susceptibility behaves like $\chi(T) \sim T^{-1+x \xi_0}$,
leading to $\alpha=x \xi_0$. Experimentally $\alpha=0.3 \div 1.2$ as
mentioned in the Introduction. 
In the case of o-TaS$_3$ we estimate $\xi_0 \simeq 150 \div 450 a_0$
in the spin sector
as mentioned previously
and $\alpha=0.3$ obtained from specific heat and susceptibility.
We deduce the concentration of
intrinsic substitutional impurities $x_{\rm int} \simeq 0.07 \div 0.2 \%$.
The upper bound is
comparable to the nominal concentration of extrinsic Nb impurities
$x_{\rm ext} \simeq 0.5 \%$
which might explain why the exponent $\alpha$ in the experiment
is the same in the  presence or absence of extrinsic impurities.

\section{Conclusions}
\label{sec:conclu}
To conclude we have investigated several factors involved
in energy relaxation in disordered CDWs and SDWs. 
A first factor is the role of commensuration. In the commensurate
case we find that the energy landscape is symmetric since the
symmetry $\varphi(x)\rightarrow-\varphi(x)$ of the CDW Hamiltonian
is preserved by the impurity potential.
As a consequence 
the two energy minima are degenerate, a property that can be
obtained by a direct solution of the sine-Gordon equation.
We have shown that
the degeneracy exists
for an arbitrary number of impurities contributing to pinning the
bisoliton. Experimentally there exists also slow relaxation in
commensurate systems, even though faster than in the incommensurate
case. The fully classical model might be too schematic since the exact
degeneracy of the classical model is lifted by quantum tunneling.
Nevertheless it succeeds to explain the $C_v\sim T^{-2}$ specific
heat and the waiting time dependence of the prefactor in the
incommensurate case.
We have proposed a model of clustering in which the pinning
energy is additive if two impurities are at a distance smaller than
the width $\xi$ of the soliton. The barrier distribution is exponential,
with therefore a power-law relaxation compatible with experiments
(see Ref.~\onlinecite{Melin-CDW}). However bisolitons can be depinned
in sequence, not simultaneously. This is described qualitatively
by the dynamical renormalization group\cite{Melin-CDW}.
Within this treatment we obtain
also a power-law relaxation but with an upper cut-off in the spectrum
of relaxation times, in agreement with experiments. The clustering
model is nevertheless useful for addressing qualitatively the waiting time
effect in the specific
ageing protocol used in experiments.

Finally we discussed another important factor: the nature of
disorder at the level of a single soliton
and the possibility of quantum interactions among solitons that
might explain the differences between 
the spin-Peierls compound Cu$_{1-x}$Zn$_x$GeO$_3$ that
does not show slow relaxation and orders antiferromagnetically,
and the spin-Peierls compound (TMTTF)$_2$PF$_6$
that shows slow relaxation without antiferromagnetic order.
For this purpose we have generalized to incommensurate systems
a model of substitutional
disorder originally introduced for spin-Peierls systems\cite{FM}.
We first treated the case of a strong pinning impurity potential and
found that 
the energy spectrum of a single quantum mechanical
soliton 
is constant as the position of the soliton varies along the
chain in the commensurate case, whereas it varies in the
incommensurate case. Interactions among solitons enhance disorder
effects in the incommensurate case. We thus find a qualitative
difference between the commensurate and incommensurate cases 
in the quantum limit. 
For substitutional disorder in the quantum limit
we find interactions among
solitons due to Friedel oscillations.
This can explain the experimentally observed power-law
contribution to the specific heat in inorganic CDWs and SDWs
in the limit of weak interchain couplings.

The final picture for organic SDWs and CDWs is
a coexistence between strong pinning and substitutional
disorder as well as a a coexistence between classical and quantum effects.
Further investigations would require
a model of collective effects interpolating between the classical
and quantum limits. 
Another ingredient that we did not discuss in detail is
metallic island formed around impurities\cite{island}
that would lead to a phenomenology close to that of substitutional disorder
because both generate states at the Fermi level for an isolated impurity,
that can interact through Friedel oscillations.

Finally,
we have developed
here the point of view of including collective effects
from the strong pinning limit.
The specific heat including quantum effects in the
weak pinning regime was discussed recently\cite{Schehr}.

\section*{Acknowledgments} 

The authors acknowledge fruitful discussions with P. Monceau.
K.B. acknowledges funding by CNRS through a temporary position
of associate researcher. The authors thank one of the referees
for useful comments on the first version of the manuscript.

\appendix

\section{Green's functions of a Charge Density Wave}
\label{app:Green}
\subsection{Canonical transformation}
We use the
Peierls Hamiltonian for charge degrees of freedom,
the electronic part of which is
given by Eq.~(\ref{eq:H-Peierls}),
where spinless fermions jump between neighboring sites on
a 1D chain. We suppose that the lattice is frozen
and therefore we do not include in the Hamiltonian the term 
due to the lattice deformations. 
The Hamiltonian (\ref{eq:H-elec}) is diagonal in terms of
the operators $\gamma_{k,R}$ and $\gamma_{k,L}$, where
R and L label right and left-moving fermions:
\begin{equation}
{\cal H} = \sum_k \left[
E_{k,R} \gamma_{k,R}^+ \gamma_{k,R}
+E_{k,L} \gamma_{k,L}^+ \gamma_{k,L} \right]
,
\end{equation}
with
\begin{eqnarray}
\gamma_{k,R}^+ &=& {\cal N}_k^{(R)} \left[
c_{k,R}^+ + {\cal B}_k^{(R)}
c_{k-2 k_F,L}^+ \right]\\
\gamma_{k,L}^+ &=& {\cal N}_k^{(L)} \left[
c_{k,L}^+ + {\cal B}_k^{(L)}
c_{k+2 k_F,R}^+ \right]
,
\end{eqnarray}
and
\begin{eqnarray}
E_{k,R} &=& 2 t \cos{(k a_0)}-
\frac{\epsilon^2}{4 t \sin{(k_F a_0)}} \frac{1}{(k-k_F)a_0}\\
E_{k,L} &=& 2 t \cos{(k a_0)}+
\frac{\epsilon^2}{4 t \sin{(k_F a_0)}} \frac{1}{(k+k_F)a_0}
.
\end{eqnarray}
The coefficients ${\cal B}_k^{(R)}$ and
${\cal B}_k^{(L)}$ are given by
\begin{eqnarray}
{\cal B}_k^{(R)} &=&-
\frac{\epsilon}{4 t} \frac{\exp{(i k_F a_0)}}{(k-k_F)a_0}\\
{\cal B}_k^{(L)} &=&
\frac{\epsilon}{4 t} \frac{\exp{(-i k_F a_0)}}{(k+k_F)a_0}
.
\end{eqnarray}
The normalization coefficients are given by
${\cal N}_k^{(R,L)}=1/\sqrt{1+|{\cal B}_k^{(R,L)}|^2}$.
The Green's functions deduced from the spectral representations
are given in Appendix~\ref{app:Green}.

\subsection{Green's functions}

The advanced Green's function defined as
$g_{y_1,y_2}(t_1,t_2) = -i \theta(t_1-t_2)
\langle \left\{ c_{y_1}(t_1), c_{y_2}(t_2) \right\} \rangle$ decomposes
into the sum of the four right (R) and left (L) combinations:
\begin{eqnarray}
g_{y_1,y_2}(t_1,t_2) = g_{y_1,y_2}^{R,R}(t_1,t_2)
+ g_{y_1,y_2}^{R,L}(t_1,t_2)\\\nonumber
+ g_{y_1,y_2}^{L,R}(t_1,t_2)
+ g_{y_1,y_2}^{L,L}(t_1,t_2)
,
\end{eqnarray}
where the ``RR'' Green's function is defined by
\begin{equation}
 g_{y_1,y_2}^{R,R}(t_1,t_2) = \sum_{k_1,k_2}
e^{i k_1 y_1} e^{-i k_2 y_2}
\langle \left\{ c_{k_1,R}(t_1), c_{k_2,R}(t_2) \right\} \rangle
,
\end{equation}
and similar expressions are obtain for the ``RL'', ``LR''
and ``LL'' Green's functions.
The spectral representation of $g_{y_1,y_2}^{R,R}$ is given by
\begin{eqnarray}
g_{y_1,y_2}^{R,R}(E) &=& \sum_k \left[ {\cal N}_k^{(R)}
\right]^2
e^{i k (y_1-y_2)} \\
&\times& \left\{
\frac{1}{E-E_{k,R}}
+ \frac{ |{\cal B}_k^{(R)}|^2}{
E-E_{k-2k_F,L}} \right\}
,
\end{eqnarray}
and similar expressions are obtained for the three other
Green's functions. After performing
the integral over wave vector in the spectral representations
we obtain
\begin{eqnarray}
&& g_{y_1,y_2}^{R,R}(E) + g_{y_1,y_2}^{L,L}(E) =
\frac{2}{3} \frac{1}{2 t \sin{k_F}} \frac{\sqrt{2}\epsilon}
{\sqrt{2 \epsilon^2-E^2}}\\\nonumber
&& \sin{\left\{\varphi + \left[ k_F-
\frac{E}{4 t \sin{k_F}} \right] R \right\}}
\exp{(-R/\xi(E))}\\\nonumber
&+&\frac{1}{3} \frac{1}{2 t \sin{k_F}}
\frac{\sqrt{2} \epsilon}{\sqrt{2 \epsilon^2-E^2}}\\\nonumber
&&\sin{\left\{\varphi-\left[ k_F+
\frac{E}{4 t \sin{(k_F a_0)}} \right] R \right\}}
\exp{(-R/\xi(E))}
,
\end{eqnarray}
where $R=y_1-y_2$ is positive and
$\xi(E)=4 t \sin{(k_F a_0)}/\sqrt{2\epsilon^2-E^2}$
is the coherence length at a finite frequency. The
sum of the ``RL'' and ``LR'' Green's functions is given by
\begin{eqnarray}
&& g_{y_1,y_2}^{R,L}(E) + g_{y_1,y_2}^{L,R}(E) =\\\nonumber
&-&\frac{\sqrt{2}}{3} \frac{1}{2 t \sin{(k_F a_0)}}
\frac{\sqrt{2}\epsilon}{\sqrt{2 \epsilon^2-E^2}}
\exp{(-R/\xi(E))}\\\nonumber
&& \left\{ \cos{\left[ k_F + 2 \varphi + \left[
k_F-\frac{E}{4 t \sin{(k_F a_0)}}\right]R 
-2 k_F y_1 \right]}\right.\\\nonumber
&-& \left.\cos{\left[ k_F-\left[ k_F-\frac{E}
{4 t \sin{(k_F a_0)}}\right] R-2 k_F y \right]}
\right\}
.
\end{eqnarray}
We obtain similar expressions for $R<0$.

\end{document}